\documentclass[9pt,twocolumn,twoside]{osajnl}
\journal{ol} % Choose journal (ao,jocn,josaa,josab,ol,optica,pr)

\setboolean{shortarticle}{true}
\usepackage{graphicx}% Include figure files
\usepackage{dcolumn}% Align table columns on decimal point
\usepackage{bm}% bold math
\newcommand{\red}[1]{{\color{black!80!black}#1}}
\usepackage{blindtext}
\usepackage{comment}
\usepackage{multirow}
\usepackage{booktabs}
\usepackage{ulem}
\usepackage{float}
\usepackage{xcolor}

\title{Thermo-optical reshaping of SHG emission from dimer all-dielectric nanoresonators}

\author[1]{Olesia Pashina}
\author[1]{Kristina Frizyuk}
\author[1]{George Zograf}
\author[1,*]{Mihail Petrov}

\affil[1]{Physics and Engineering Department, ITMO University, St. Petersburg, Russia, 197101}
\affil[*]{Corresponding author: m.petrov@metalab.ifmo.ru}

\begin{abstract}
In recent years resonant semiconductor and all-dielectric nanophotonics offered a lot of possibilities for thermally-induced light manipulation at the nanoscale. Owing to high-quality resonant states, such nanostructures allow for efficient light-to-heat conversion supported by various temperature detection approaches based on thermally sensitive intrinsic optical responses (photoluminescence, Raman scattering, thermorefraction etc.). In this work, we study theoretically a phenomenon of the photothermal reshaping of the radiation pattern of second-harmonic generation (SHG) that occurs in resonant all-dielectric systems. In the suggested geometry near-IR pulsed laser is utilized for excitation of SHG while simultaneous continuous wave visible laser heats the structure. The thermooptical switching of the resonant optical states in the nanostructures governs the reconfiguration of the emission pattern, without significant loss in the magnitude of the SHG. We believe, that our findings will pave the way for \red{subwavelength-size} near-IR thermally switchable nonlinear optical devices. 
\end{abstract}

\setboolean{displaycopyright}{true}

\begin{document}

\maketitle

After the first observation of the phenomenon of optical harmonics generation \cite{Franken1961-GenerationofOptical}, such nonlinear optical effect proved its high efficiency in crystals and waveguides  \cite{Yamada1993-xn--Firstorder-ut6exn--quasipha-4m3d}, becoming a commercially available technology for optical microscopy \cite{Campagnola2011-SecondHarmonicGener,Campagnola2003-Second-harmonicimagi}, or laser sources and even green-light laser pointers. While high efficiencies of SHG in noncentrosymmetric bulk materials could be achieved by a phase-matching \cite{Fejer1992Nov}, going down to subwavelength-scale particles and retaining high efficiency was quite a challenge. Different plasmonic~\cite{Celebrano2015-Modematchinginmult, Butet2015-OpticalSecondHarmon,Mochizuki2020-SecondHarmonicGener,Finazzi2007-Selectionrulesfors,Noor2020-Mode-MatchingEnhance,Elkabetz2021-Optimizationofsecon,Nadolski2020-AdverseRoleofShape,Gurdal2020-Enhancementofthese}, all-dielectric~\cite{Makarov2017-EfficientSecond-Harm,Smirnova2016-Multipolarnonlinear,Jiang2020-Strongsecond-harmoni,Gigli2020-Quasinormal-ModeNon,Xu2018-Highly-EfficientLong,Zeng2020-Enhancedsecond-harmo,volkovskaya2020multipolar,Cambiasso2017-BridgingtheGapbetw,Kroychuk2020-EnhancedNonlinearLi}, and even hybrid (metal-dielectric)~\cite{Grinblat2014Nov, Timpu2017-EnhancedSecond-Harmo,Yang2018-Efficientsecondharm,Chervinskii2018-NonresonantLocalFie,Scherbak2018-UnderstandingtheSec,Hu2020-Second-harmonicgener,Linnenbank2016-Secondharmonicgener} designs were proposed, however, the conversion of the pump to second harmonic (SH) signal was still of the order of $10^{-5}-10^{-4}$. \red{Moreover, very recent attractive family of materials for nanophotonics community - transition metal dichalcogenides (TMDCs) - both monolayers~\cite{wang2015giant} and nanoresonators~\cite{busschaert2020transition} demonstrated its potential in nonlinear nanophotonics. Even though TMDC monolayers mostly posses a very high nonlinear susceptibility, the bulk second-order nonlinearity tensor is zero due to crystalline symmetry for most of the materials~\cite{boyd2020nonlinear}. There are other approaches for boosting SH conversion based on bound states in the continuum (BIC) or quasi-BIC modes in resonant structures (arrays or metasurfaces) reaching $10^{-2}-10^{-1}$, but these structures are no longer comparable to the wavelength scale~\cite{kang2021efficient,ning2021ultimate,han2021significantly}. Moreover, high-Q states in single cylindrical  subwavelength resonators supported $10^{-1}$ SH conversion~\cite{Koshelev2020Jan}, however tunability of modes and radiation pattern still remains a challenge.}

On the other hand, such nanoantennas proved their ability for strong optical heating and thermometry due to resonantly enhanced optical absorption~\cite{Zograf2017May,Zograf2021,Yazdanfar2020-Modalcontrolofther}. Owing to strong thermooptical effects~\cite{Tsoulos2020Sep,Duh2020Aug}  all-dielectric structures already demonstrated their proficiency in modulation of linear properties of propagating \cite{Rahmani2017Aug,Kamali2019Apr} and scattered  \cite{Zhang2020Jun} light.  Moreover, the possibility of all-optical thermal modulation of the SHG emission intensity in resonant semiconductor nanostructures has been reported~\cite{Celebrano2021} recently, which makes  it a an accessible experimental technique.

%====================FIGURE1==============================
\begin{figure}[t!]
  \includegraphics[width=0.99\linewidth]{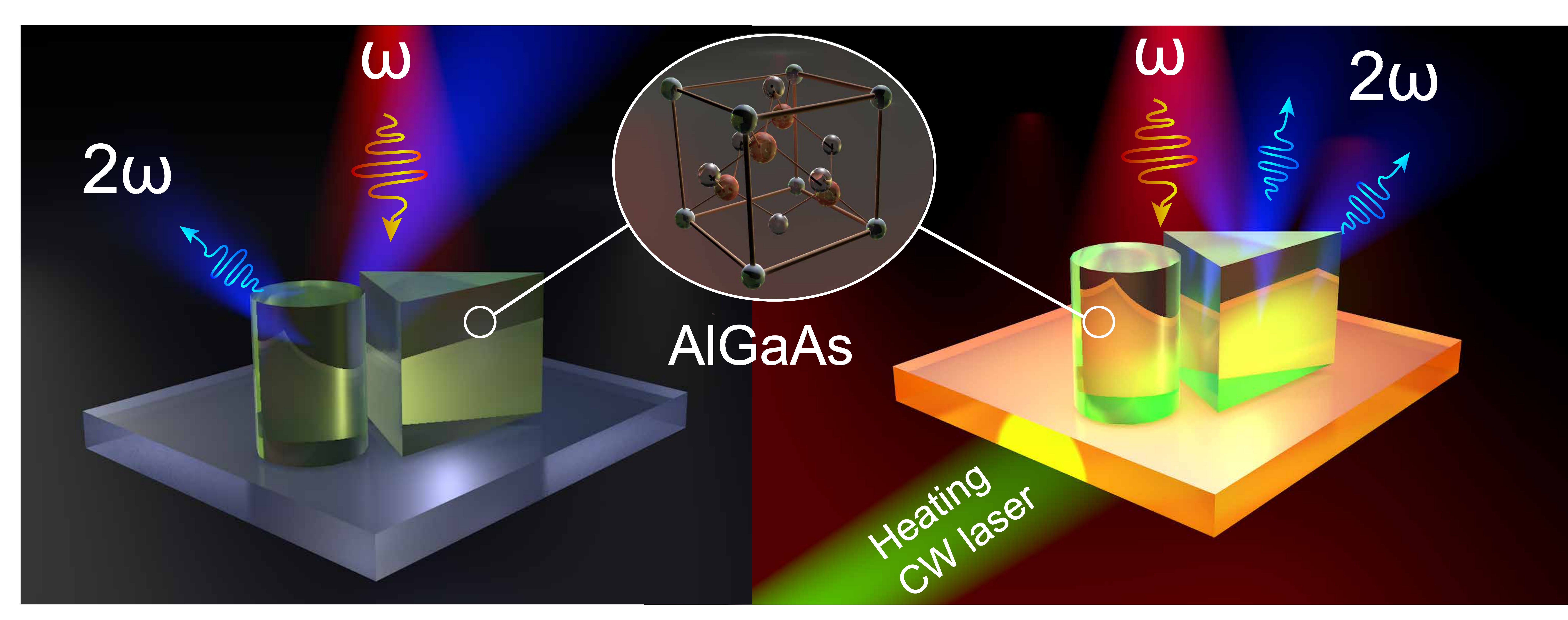}% Here is how to import EPS art
\caption{\textbf{Schematic of the  concept.}  Continuous wave heating of a dimer nanostructure steers the SHG emission  far-field pattern. The dimer consists of a triangular prism and a cylinder of  Al$_{x}$Ga$_{1-x}$As located on a glass substrate.}
\label{fig:1}
\end{figure}
%====================FIGURE1==============================
 In this work, we propose $\textit{in situ}$ thermooptical tuning of SHG pattern in  bi-resonant dimer nanoantenna, \red{consisting of two nanoresonators, thereby making the whole nanoantenna of subwavelength size} (see Fig.~\ref{fig:1}). Being a highly important property  of nanoscale light emission, the directivity pattern of the SHG was well studied for single resonant nanoantennas \cite{Carletti2016Aug,Gigli2021Feb}. However, under the intense continuous wave (CW) laser excitation \cite{Celebrano2021, Zhang2020Jun} the spectral position of resonances can shift due to thermooptical effects leading to reconfiguration of the emission pattern. In this view, we \red{ advance the approach suggested in \cite{Celebrano2021} proposing two closely located nanostructures forming a dimer nanoantenna. Different particles shape and slight spectral detuning between the resonances allow for reconfiguration of the nonlinear emission pattern with the temperature increase as schematically shown in Fig.~\ref{fig:1}.  Such an approach, when each single element is tuned at particular temperature regime, can be utilized for desingning  thermally stable or tunable nonlinear optical systems.}
% missing refererences 
% \cite{frizyuk2019JOSA,frizyuk2019second,timofeeva2018anapoles,shcherbakov2014enhanced,shorokhov2016multifold, liu2016resonantly,renaut2019reshaping,xu2019forward,Carletti2016Aug}
%-------------------------FIGURE2---------------
\begin{figure}[t]
\includegraphics[width=.99\linewidth]{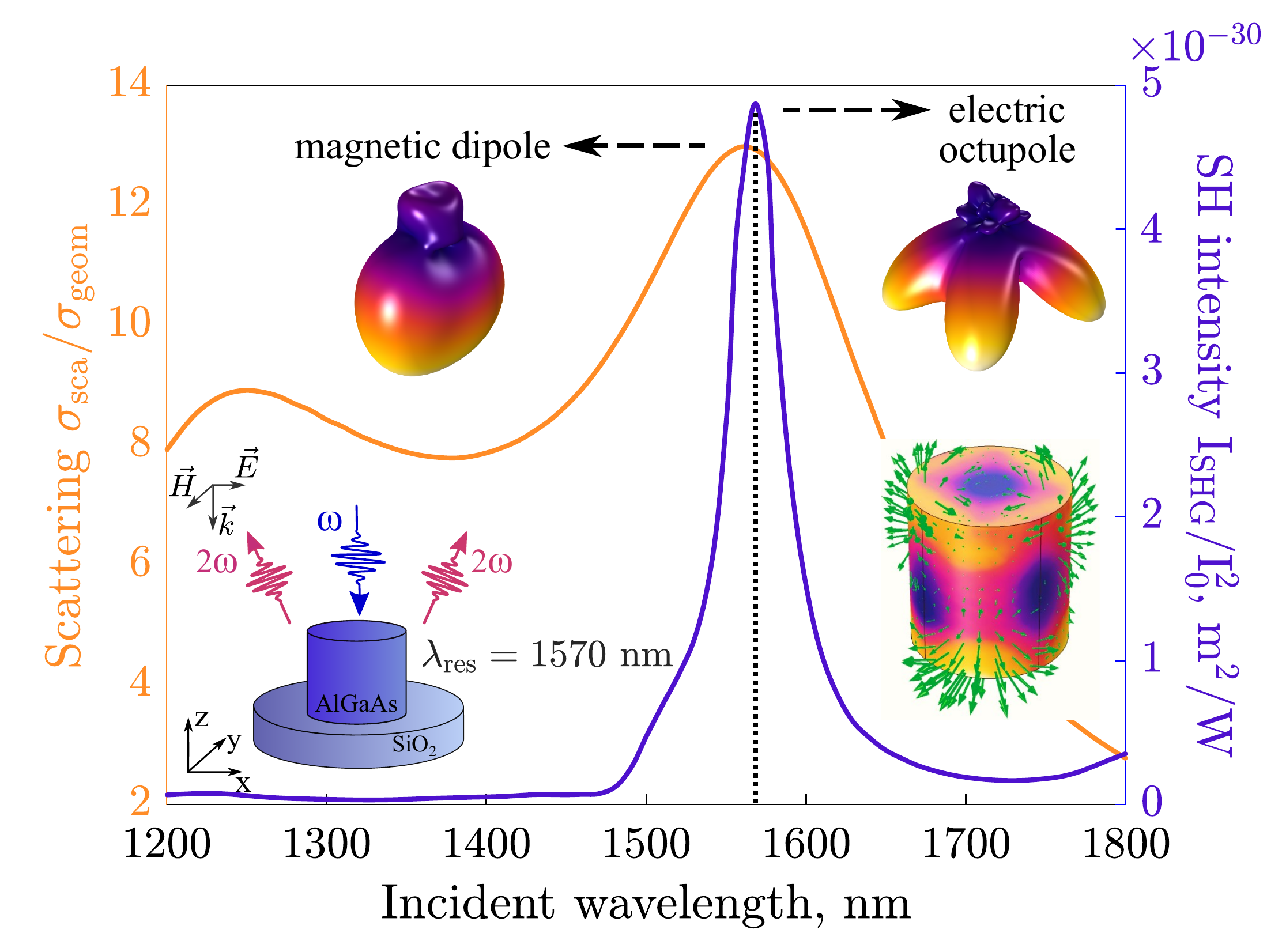}% Here is how to import EPS art
\caption{\textbf{Resonant properties of a single nanocylinder for the incident and second harmonic frequencies.} 
A nanodisk with a diameter $\mathrm{d} = 386 \: \mathrm{nm}$ and a height $\mathrm{h} = 400 \: \mathrm{nm}$, placed on the glass substrate is excited with normally incident plane wave. The orange color shows the magnetic dipole resonance of linear scattering, the violet color refers to the electric octupole resonance of the SHG. The SHG intensity was normalized to the square of the incident intensity $\mathrm{I_{SHG}/I_{0}^2}$, and the scattering cross-section was normalized to the geometric section of the particle ${\sigma_{sca}/(\pi R^2))}$, where $R$ is the radius of the cylinder. Far-field radiation patterns are shown for both linear scattering and SHG. \red {In the lower left corner the polarization arrows and the magnitude of the Electric field on the surface of the nanoparticle are shown in green.}}\label{fig:2}
\end{figure}
%---------FIGURE2-------------------------

We rely on semiconductor solid solution of  Al$_x$Ga$_{1-x}$As which is an efficient nonlinear material with large quadratic nonlinear optical coefficient $d\approx 100$ pm/V at wavelengths of about 1550 nm ~\cite{Shoji1997Sep, Ohashi1993Jul}. \red{ This coefficient is related to the second-order nonlinear susceptibility as $d=\frac{1}{2} \chi^{(2)}$~\cite{boyd2020nonlinear} and determines the nonlinear polarization for the material under consideration according to the equation: $P_{i}=2\varepsilon_{0} d\sum_{j k} \delta_{i j k} E_{j} E_{k}$, where $\delta_{i j k}$ vanishes, if any of the two indices i, j, k coincide and $\delta_{i j k}=1$ otherwise.} For  further  simulations, we use particular compound $x=0.198$ which properties are available elsewhere \cite{aspnes1986optical,Gehrsitz2000Jun,Adachi1985Aug,afromowitz1973thermal}. We start with analyzing the nonlinear and thermooptical response of a single cylindrical nanostructure. The parameters of a cylinder structure can be optimized in a way to achieve double resonant condition both at the fundamental and SH wavelengths. Indeed, the resonant mode excitation at the fundamental wavelength is vital for achieving  the enhancement of the SHG signal~\cite{frizyuk2019second,Linnenbank2016-Secondharmonicgener,Zeng2018-EnhancedSecondHarmo,Thyagarajan2012-Enhancedsecond-harmo,Park2012-Doublyresonantmetal,Wang2020Sep}. On the other hand, the resonance at second harmonic emission wavelength $\lambda_{\mathrm{pump}}/2$ provides additional  enhancement of the SHG signal and, more importantly, governs  the SH emission pattern ~\cite{Carletti2016Aug, frizyuk2019second}). 
Fig.~\ref{fig:2} depicts the elastic light scattering properties and SHG efficiency of a single AlGaAs nanodisk optimized for pumping in the infrared region 1550 - 1600 nm typical for  commercially available pulsed-laser systems. At roughly 1570 nm the nanodisk of 386 nm diameter and 400 nm height possesses resonant  magnetic dipole response, allowing for enhancement of the electromagnetic field inside the nanodisks volume. \red{ At the same time, at the SH  wavelength the  sharp resonant enhancement of the emission is observed.  The distribution of the SH electric field on the nanoparticle surface shown in Fig.~\ref{fig:2}, as well as the corresponding radiation pattern with four pronounced lobes due to the substrate leakage, are characteristic of the octupole mode with the axial number m=2 \cite{Gladyshev2020}. }  The pronounced lobes in the lower direction correspond to the preferential substrate emission. 

%----------------------FIGURE 3--------------
\begin{figure}[ht!]
\includegraphics[width=.99\linewidth]{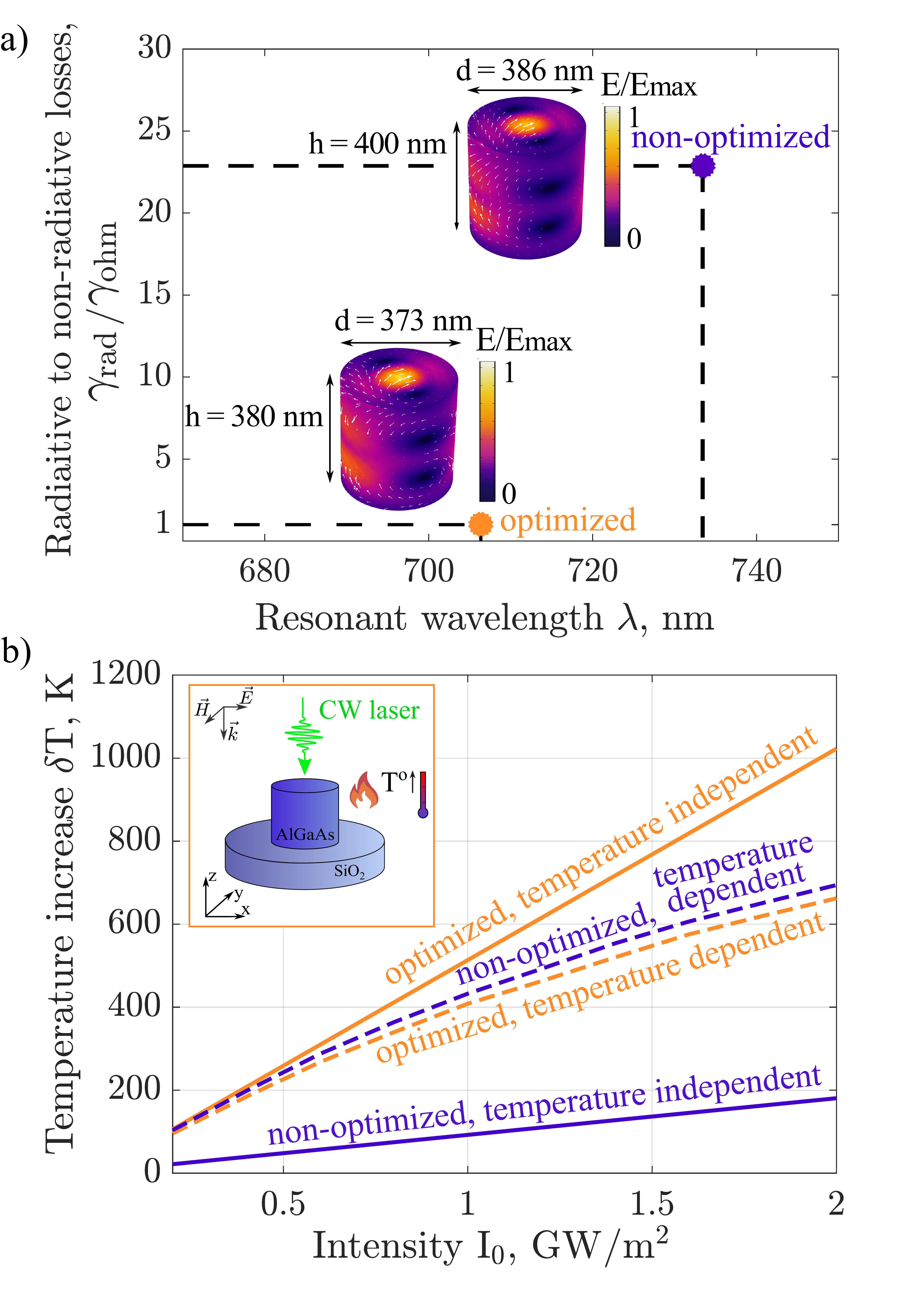}% Here is how to import EPS art
\caption{{\textbf{Critical-coupling heating regime and self-consistent optical heating of a single nanoparticle.} a) The ratio of radiation and ohmic losses $\gamma_{\mathrm{rad}}/\gamma_{\mathrm{Ohm}}$ for the eigenmode of cylinders with optimized ($\mathrm{h} = 380 \: \mathrm{nm}$ and $\mathrm{d} = 373 \: \mathrm{nm}$) and non-optimized ($\mathrm{h} = 400 \: \mathrm{nm}$ and $\mathrm{d} = 386 \: \mathrm{nm}$) size. The optimized case provides the critical-coupling heating regime, corresponding to the equality of ohmic and radiation losses $\gamma_{\mathrm{rad}}=\gamma_{\mathrm{Ohm}}$. b) The temperature increase $\delta \mathrm{T}$ dependence on the CW laser intensity at optimized (orange lines) and non-optimized (violet lines) cylinder parameters. The solid lines show are obtained with simulations  without taking into account temperature-dependent optical constants, while the dashed lines show the self-consistent optical heating modeling.}} \label{fig:3}
\end{figure}
%----------------------FIGURE 3--------------

%====================================FIGURE4==========================
\begin{figure*}[t!]
\includegraphics[width=1.9\columnwidth]{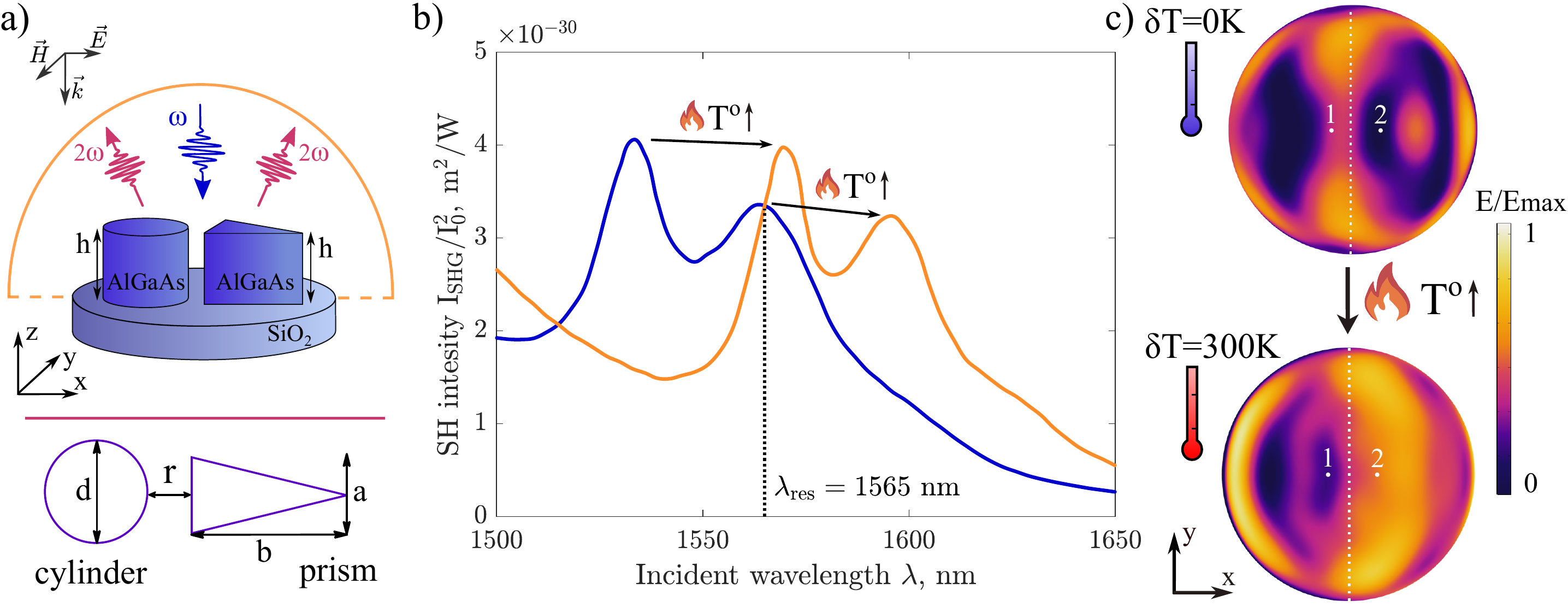}% Here is how to import EPS art
\caption{\label{fig:4} {\textbf{Thermally tunable bi-resonant dimer radiation patterns.} a) Schematic representation of a dimer structure consisting of of a cylinder and a prism with the following geometric parameters $\mathrm{h} = 400\: \mathrm{nm}$, $\mathrm{d} = 386\: \mathrm{nm}$,
$\mathrm{a} = 350 \:\mathrm{nm}$, $\mathrm{b} = 460\: \mathrm{nm}$, $\mathrm{r} = 200\: \mathrm{nm}$. The orange semicircle represents the hemisphere of the second harmonic signal detection from the dimer.
 b) The dependence of SHG intensity normalized to the square of the incident intensity $\mathrm{I_{SHG}/I_{0}^2}$, \red{($\mathrm{I_0}=10^{13}\:\mathrm{W/m^2}$\cite{liu2016resonantly})} at two temperature values for a dimer consisting of a prism and a cylinder. The blue line corresponds to 25$^o$ C and the orange line is for the case of the elevated temperature of $ 325 ^ o $ C. The fundamental wavelength is 1565 nm and is marked with the vertical black line.
c) Top view of the SHG radiation detected on the upper hemisphere of a dimer consisting of a cylinder and a prism at a wavelength of 1565 nm at room $\delta \mathrm{T} = 0 \: \mathrm{K}$ and elevated $\delta \mathrm{T} = 300 \: \mathrm{K}$ temperature. The electric field E is polarized along the x-axis. The radiation patterns at room temperature and at elevated temperature have the same colorscale. \red{The dominant directions of strong switching for SH intensities during heating are marked with white points 1 and 2, and the detection hemisphere is divided into left and right parts by a white dotted line.} }}
\end{figure*}
%====================================FIGURE4==================

At the next step, we pay attention to thermooptical response of the considered structures. In accordance with the recent experimental successes~\cite{Celebrano2021}, we consider the normal incidence of continuous wave laser beam that heats the nanostructure.  \red{The laser-assisted optical heating has a big advantage of tunable and local heating of the nanostructure~\cite{aouassa2017temperature} rather than changing the temperature of the whole setup.} Owing to the strong spectral dependence of the material losses in the semiconductor structures, the proper choice of the heating laser wavelength and excitation conditions is  crucial on the way of achieving  strong thermooptical response. It is well known that the radiative ($\gamma_{\mathrm{rad}}$) and non-radiative ($\gamma_{\mathrm{Ohm}}$) losses rate of particular mode should be balanced~\cite{Zograf2017May,miroshnichenko2018ultimate,Zograf2021}  providing  the optimal heating condition. One can easily balance  in semiconductors in the region of strong interband absorption, which in the case of crystalline silicon occurs around 700 nm region.  Having fixed the ratio of the height and diameter of the nanocylinder $ \mathrm{h / d} $, we slightly scale the geometric parameters of the resonator. As a result, the eigenmode is spectrally shifted and, while the radiative losses do not change significantly, the non-raidiative losses are fastly changing due to their strong spectral dependence (see Fig.~\ref{fig:3} (a)). Therefore,  the ratio $\gamma_{\mathrm{rad}}$/$\gamma_{\mathrm{Ohm}}$ takes different values reaching the optimal condition for disks of 373 nm diameter. \red{Worth mentioning, that we are neglecting the pulsed laser heating assuming that the pulse intensity is short enough and the repetition rate is low ~\cite{pashina2021ultrafast}}.

The thermooptical simulations performed with commercially available COMSOL Multyphiscs software show that the heating efficiency of the optimized structure (orange line)  is much higher than that of the non-optimized (blue) which is a clear sign of critical coupling regime. \red{The intensity range was chosen well below the damage threshold\cite{liu2016resonantly}}. However, these data were obtained for temperature-{\it independent} material losses without taking into account the thermooptical effects. At the same time,  intense optical heating induces a change in the refractive index of the nanoobject material according to a linear law $ n(T) = n_0 + \alpha T$, where $\alpha$ is the thermorefractive coefficient. The typical value of  thermorefractive coefficients for AlAs  $\alpha=1.25\cdot10^{-4} {K}^{-1}$, at 750 nm wavelength~\cite{Gehrsitz2000Jun}. The imaginary part of the refractive index also has strong dependence on temperature $k(T)$. However, these parameters also depend both on the wavelength and particular compound of solid mixture $x$, and they were interpolated from the data available for AlAs and  GaAs~\cite{Yao1991Sep, Yao1992} taken in the appropriate proportion corresponding to solid mixture Al$_x$Ga$_{1-x}$As for \red{$x=0.198$}. Surprisingly, including the thermorefractive  effect $n(T), k(T)$ into the simulations  leads to rearrangement of the losses channels in the nanoresonator:  absorption in initially non-optimized geometry fastly increases with the temperature and to match the radiative losses and locking the system in the balance losses condition. 
Now, the originally optimized and non-optimized disks have very similar intensity dependence (Fig.~\ref{fig:3}b), and non-optimized geometry shows even slightly stronger thermooptical response, which we utilized for further simulations. \red{The optimal heating conditions for temperature dependent optical parameters is a matter of independent study and will be reported in the following works.} \red{The figure depicts a steady-state solution of CW laser heating of single nanodisk on a glass substrate. We assume the establishment of average temperature within one nanoparticle and the dimer as whole, since the thermal conductivity of our semiconductor nanostructure $\kappa_{AlGaAs}$ = 22.75 W m/K is much higher than the thermal conductivity of air $\kappa_{air}$= 0.026 W m/K and glass $\kappa_{SiO_2}$= 0.8 W m/K, adjacent to the surface of nanoparticles. In this regard, homogeneous thermal distribution within our nanodimer becomes a fair approximation, because an extremely small part of the heat leaks into the substrate or into the air~\cite{Baffou2017, Zograf2021}.}

\begin{figure}[t!]
\includegraphics[width=1\columnwidth]{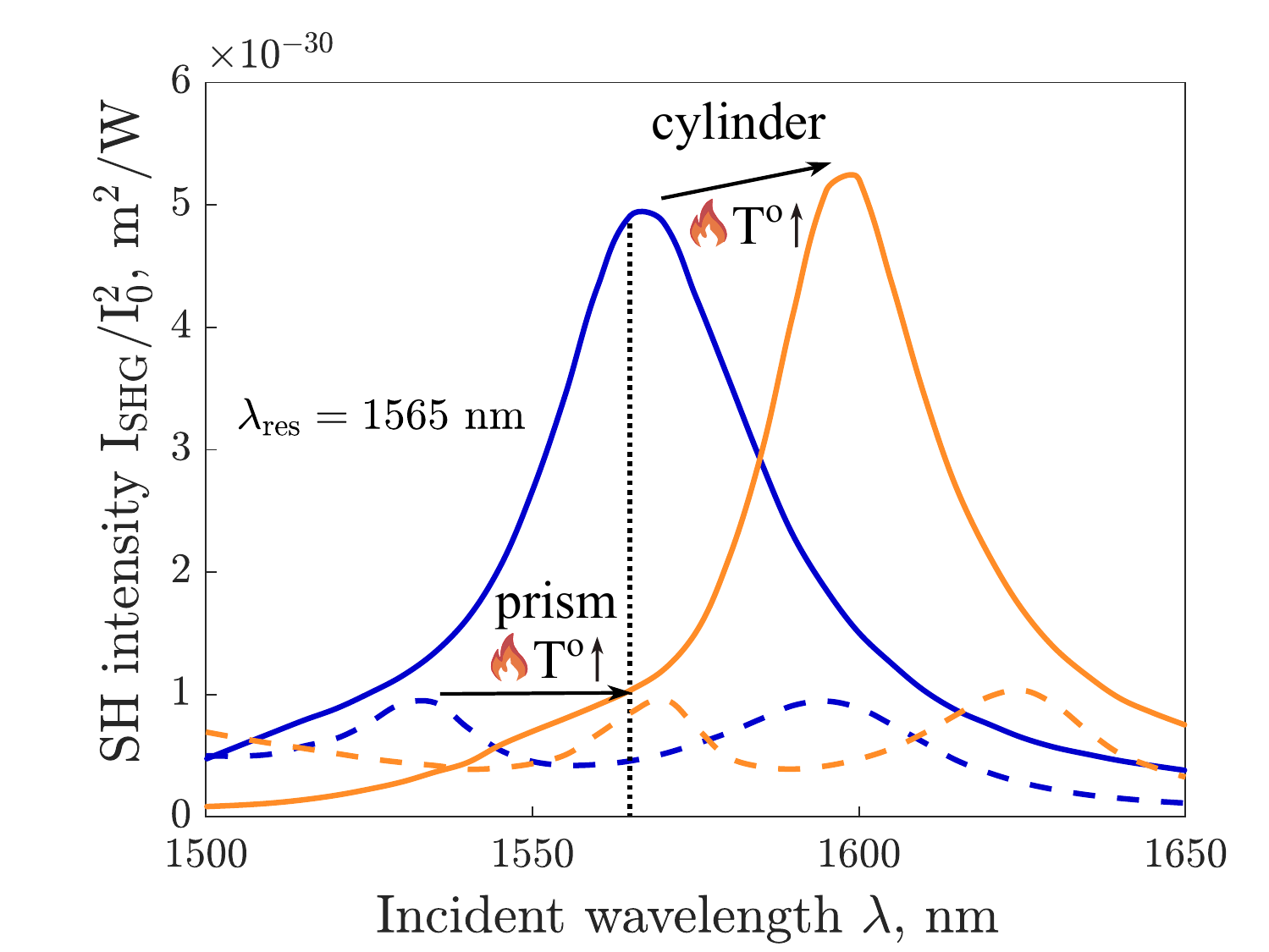}% Here is how to import EPS art
\caption{\label{fig:5}\red{ {\textbf{Second harmonic generation from individual dimer components.} 
The dependence of SHG intensity normalized to the square of the incident intensity $\mathrm{I_{SHG}/I_{0}^2}$, where $\mathrm{I_0}=10^{13}\:\mathrm{W/m^2}$, at two temperature values separately from the cylinder (solid lines) and prism (dotted lines) that make up the dimer. The blue lines corresponds to 25$^o$ C and the orange line is for the case of the elevated temperature of $ 325 ^ o $ C. The fundamental wavelength is 1565 nm and is marked with the vertical black line.
}}}
\end{figure}

Thus, one of the main thermooptical effects observed in the nanoresonator structures is related to the spectral  shift of the resonant spectral line due to thermorefractive effect. Inevitably, that leads to taking the system out of the resonant condition both at fundamental and SH wavelengths. As soon as the dominant multipole contribution is suppressed the other non-resonant multipoles start to influence the nonlinear response leading to reconfiguration of the far-field radiation pattern.  In this regard, we consider a nanoantenna structure  compound of two elements (dimer structure), which resonances are detuned from each other for the value corespondent to the value of thermal drift of the resonance. Thus, one of the elements of the dimer will be in the resonant state in the cold regime, while another at the elevated temperature. \red{Speaking of experimental realization, it was already experimentally shown that SH emission can be thermally detuned and detected~\cite{Celebrano2021}, moreover, asymmetric nanoresonators can be fabricated as well~\cite{Gigli2021Feb}.} In order to achieve reconfiguration and switching  of the SH far-field  pattern, one should also choose the different shapes and symmetries of "on" and "off" structures. In this regard, we have chosen a cylinder-prism structure. One may fairly ask a question why should it be particularly a prism. The reason for that lies in the symmetry governed nonlinear response of nanoantenna structures:  by lowering the symmetry of the structure one enables the  excitation of the modes, which are prohibited in the case of a disk \cite{frizyuk2019JOSA}. In this regard, the sphere, cube, or cone have too high symmetries, while prism has well-pronounced resonances, low symmetry, and are also relatively easy to fabricate with planar technique. The cylinder-prism structure  under the consideration is shown in Fig.~\ref{fig:4} (a) along with the geometry of the excitation and collection aperture (shown in yellow circle). In the proposed structure\red{, in accordance with Fig.~\ref{fig:5},} the cylinder \red{possesses the SH resonance at the fundamental wavelength and therefore} is in the "on"-regime at room temperature, while prism gets into the resonant state at the elevated temperature. The simulated SH spectrum \red{from dimer structure} is shown in  Fig.~\ref{fig:4} (b) at room temperature (blue line) and at the elevated temperature of 325$^o$ C (orange line). One can see that the SH spectrum has a two-peak structure, which are separated by a spectral gap of 50 nm and inherent the spectral detuning between the resonances of cylinder and prism structures. Once designed to have the long-wavelength SHG resonance at 1565 nm at room temperature, the short-wavelength peak gets into the resonant state at elevated temperature. This is accompanied by strong reshaping of the SH far-field patterned which is governed by different responses of the cylinder and prism structures. \red{For the two most dominant directions, we obtained the ratio of the SH intensities $\mathrm{
\left.I\right|_{325^o C}/\left.I\right|_{25^o C}}=0.5$ and  $\mathrm{
\left.I\right|_{325^o C}/\left.I\right|_{25^o C}}=23.7$ for points 1 and 2 respectively upon switching from room to elevated temperatures. The SHG value integrated over the right hemisphere from the heated dimer was approximately 2.8 times higher than the signal from the room temperature dimer.} While at low-temperature the SH pattern was almost symmetric \red{because of  the main contribution to the dimer SH resonance from the included "on"-cylinder,} it becomes strongly asymmetric at a higher temperature \red{ due to the influence of the commensurate radiations from both dimer elements}.

\red{Finally, we theoretically proposed an approach to control the directivity of SH emission from nanoantenna structure with help of thermooptical effect, which efficiency can be enhanced under proper designing of subwavelength high-Q systems.    }
%====================================FIGURE5==========================
%====================================FIGURE5==================
\begin{backmatter}

\bmsection{Acknowledgments} The work was supported by the Grant Of President of Russian Federation (grant no. MK-2360.2020.2) and Russian Foundation of Basic Research (grant no. 20-32-90238). M.P. acknowledges support from the federal academic leadership program Priority 2030.

\bmsection{Disclosures} The authors declare no conflicts of interest.

\end{backmatter}

% Bibliography
\bibliography{apssamp}

%\bibliographyfullrefs{apssamp}

\end{document}